\documentclass{aa}

\usepackage{graphicx}
\usepackage{natbib}
\usepackage{txfonts}

\begin{document}

\title {The Effects of radial inflow of gas and galactic fountains on the chemical evolution of M31 }

\author {E. Spitoni\inst{1}
\thanks {email to: spitoni@galaxy.lca.uevora.pt}
\and  F. Matteucci\inst{2}
\and M. M. Marcon-Uchida\inst{3} }

\institute{ Department of Mathematics,  University of \'Evora, R. Rom\~ao Ramalho 59, 7000 \'Evora, Portugal
\and  I.N.A.F. Osservatorio Astronomico di Trieste, via G.B. Tiepolo 11, I-34131, Italy \and N\'ucleo de Astrof\'isica Te\'orica, Universidade Cruzeiro do Sul, Rua Galv\~ao Bueno 868, Liberdade,
 01506-000, S\~ao Paulo - SP, Brazil
}

\date{Received xxxx / Accepted xxxx}

\abstract {Galactic fountains and radial gas flows are very important
  ingredients in modeling the chemical evolution of galactic disks.}
          {Our aim here is to study the effects of galactic fountains
            and radial gas flows in the chemical evolution of the disk
            of Andromeda (M31) galaxy.}  {We adopt a ballistic method to
            study the effects of galactic fountains on the chemical
            enrichment of the M31 disk. In particular, we study the
            effects of the landing coordinate of the fountains and the
            time delay in the pollution of the interstellar gas. To
            study the effects of radial flows we adopt a very detailed
            chemical evolution model. Our aim is to study the formation of
            abundance gradients along the M31 disk and also compare our
            results with the Milky Way.}  {We find that the landing
            coordinate for the fountains in M31 is no more than 1 kpc
            from the starting point, thus producing negligible effect
            on the chemical evolution of the disk. We find that the
            delay time in the enrichment process due to fountains is
            no longer than 100 Myr and this timescale also
            produces negligible effects
            on the results. Then, we compute the chemical evolution of
            the M31 disk with radial gas flows produced by the infall
            of extragalactic material and fountains.  We find that a
            moderate inside-out formation of the disk coupled with
            radial flows of variable speed can very well reproduce the
            observed gradient. We discuss also the effects of other
            parameters such a threshold in the gas density for star
            formation and an efficiency of star formation varying with
            the galactic radius.}  {We conclude that galactic
            fountains do not affect the chemical evolution of the M31
            disk. The inclusion of radial gas flows together with an
            inside-out formation of the disk, produces a very good agreement
            with observations. On the other hand, if radial flows are
            not considered, one should assume a threshold in the star
            formation and a variable star formation efficiency,
            besides the inside-out formation to reproduce the data. We
            conclude that the most important physical processes in
            creating disk gradients are the inside-out formation
            and the radial gas
            flows. More data on abundance gradients both locally and
            at high redshift are necessary to confirm this
            conclusion.}


\titlerunning{M31 chemical evolution}
\authorrunning{Spitoni et al.}
\maketitle

\section{Introduction}
Galactic chemical evolution predicts how chemical elements are formed
and distributed in galaxies. Galactic chemical evolution models follow
the evolution of chemical abundances in the interstellar medium (ISM)
is space and time. Chemical evolution is determined by the history of
star formation, the stellar yields and gas flows. In particular, most
of chemical evolution models deal with infall and outflow of gas in
galaxies but very few models have taken into account the effects of
radial flows and even less the effects of galactic fountains on the
chemical evolution.  Galactic fountains are created by supernova (SN)
explosions in the disk of a galaxy: the gas ejected by multiple SN
explosions occurring in OB associations reaches a certain height above
the galactic plane and then, due to the potential well of the galaxy,
it falls back onto disk. Bregman (1980) modeled first this process
assuming that the gas falls back ballistically.  Spitoni et
al. (2008;2009) followed this approach and computed the effects of
fountains on the chemical evolution of the Galactic disk and concluded
that they are negligible; in fact, the gas ejected from the disk is
likely to land very close to the place where it escaped. Moreover, the
time delay with which the enriched supernova material is coming back
into the ISM ($\sim 100$ Myr) does not influence substantially the
chemical enrichment process.  Gas infall is an important ingredient in
the build -up of galactic disks and it produces radial gas flows, as
shown first by Mayor \& Vigroux (1981). In fact, the infalling gas has
a lower angular momentum than the circular motions in the disk, and
mixing with the gas in the disc induces a net radial inflow.  Lacey \&
Fall (1985) computed a chemical model with radial flows and estimated
that the gas inflow velocity is up to a few km s$^{-1}$.  Later, Goetz
\& K\"oppen (1992) studied numerical and analytical models including
radial flows. Chemical models with radial flows were studied more
recently by Portinari and Chiosi (2000), Sch\"onrich \& Binney (2009)
and Spitoni \& Matteucci (2011), among others. In particular, Spitoni
\& Matteucci (2011) concluded that models assuming no inside-out
formation of the Galactic disk need the presence of radial flows to
explain the existence of abundance gradients. On the other hand, an
inside-out formation of the disk coupled with a threshold in the star
formation could equally reproduce the observed gradients although less
steep than in the case of radial flows. Since the inside-out formation
of disks seems to be observed at high redshift (Munoz-Mateos et
al. 2007), the more reasonable conclusion was that both the inside-out
process and radial flows should be at work in the Milky Way.
 
Here, we plan to compute the chemical evolution of the
disk of M31 including both the effects of galactic fountains and radial gas
flows. We will start by adopting a model of chemical evolution of M31
developed by Marcon-Uchida et al. (2010) including inside-out
formation of the disk but no fountains nor radial flows.
The paper is organized as follows: in Section 2 we describe the model
of Spitoni et al. (2008) used to compute the fountains in M31 and in
Section 3 the results of this model are described. In Section 4 the
Marcon-Uchida et al. (2010) model plus the implementation of radial
flows in it are described. In section 5 the results for the abundance
gradients in M31 are shown. In section 6 some conclusions are drawn.

\section {The Galactic fountain model of Spitoni et al. (2008)}
In order to study galactic fountains, Spitoni et al. (2008) followed the 
evolution
of a superbubble driven by supernova explosions in the Galactic disk.
They described the superbubble evolution using
the Kompaneets (1960) approximation. Here we do not enter into the
details of this model and we just recall that Kompaneets (1960) founds
analytical expressions for the shape of the bubble during its
expansion in an exponential atmosphere with density:

\begin{equation}
\rho(z)=\rho_{0}exp(-z/H),
\label{rhozp}
\end{equation}
where $\rho_{0}$ and $H$ are the disk density and scale height, respectively.

In Spitoni et al. (2008) we showed that  the total time  necessary for the growth of instabilities and for the fragmentation of the superbubble in terms of $\rho_0$,  $H$,  and the luminosity of the system $L_0$ is:
\begin{equation}
t_{final}=4.37\times \left(\frac{\rho_{0} H^{5}}{L_{0}}\right)^{1/3}.
\label{final}
\end{equation}

Once the top of the supershell reaches the height above the galactic
plane related to the time $t_{final}$ (see eq. \ref{final}), the thin
shell can leave the stellar disk and move towards the extra-planar
gas.
 Ballistic models describe the gas as an inhomogeneous collection
of clouds, subject only to the gravitational potential of the Galaxy.
 The way we consider the galactic fountain for M31
is identical to the model described in Spitoni et al. (2008),
therefore we address the reader to that paper for all the details.

\subsection{ The galactic potential and the OB associations  of M31}
\label{galaxy}
The potential well of M31 is assumed to be the sum of three components
as suggested by Howley et al. (2008): a dark matter halo, a bulge and
a disk.  The dark matter halo gravitational potential is assumed to
follow the Navarro, Frenk and White (1996) profile:

\begin{equation}
\Phi_{h}(r) = -4\pi G\delta_{c}\rho_{c}r_{h}^{2} \left( \frac{r_{h}}{r} \right) \ln \left[ \frac{r+r_{h}}{r_h}\right] 
\label{eq:NFW}
\end{equation}
where $\delta_c$ is a dimensionless density parameter, $\rho_{c}$ is
today's critical density with Hubble constant $h=0.71$ in units of 100
km/s/Mpc, and $r_{h}$ is the halo scale radius (Navarro et al. 1996).
The bulge gravitational potential is given by (Hernquist 1990):

\begin{equation}
\Phi_{b}(r) = -\frac{GM_{b}}{a_{0} + r}
\label{eq:hernquist}
\end{equation}

where $a_{0}$ is the scale radius and $M_{b}$ is the bulge mass.

 For the disk potential we have chosen the axisymmetrical Miyamoto \&
 Nagai (1975) model which provides results that are comparable to
 those from the exponential disk used by Geehan et al. (2006). In
 cylindrical coordinates $(R,z)$ can be written as:

\begin{equation}
\Phi_{d}(R,z) = \frac{-GM_{d}}{\sqrt{R^{2}+(R_{d}+\sqrt{z^{2}+b^{2}})^{2}}},
\label{eq:miyamoto}
\end{equation}
where Rd is the disk scale length and b is the vertical scale factor.

We use as done in Howley et al. (2008) the ``Best-fit Model'' values
derived by Geehan et al. (2006) to describe the various parameters of M31, with the sole
  exception of $b$, the vertical scale factor, which was not a
  reported parameter.  For $b$ we use the vertical scale height of the
  dust at a value of 0.1 kpc (Hatano et al. 1997). The values reported by
  Geehan et al. (2006) include the bulge mass, with $M_{b}=3.3 \times
  10^{10}$ M$_{\odot}$, the bulge scale factor, with $a_{0}=r_{b}=0.61$ kpc,
  the disk central surface density, with $\Sigma_0=4.6 \times
  10^{8}$  M$_{\odot}$ kpc$^{-2}$, the disk scale radius, with $R_d=5.4$ kpc,
  the halo scale radius, with $r_h=8.18$ kpc.

 For the interstellar medium (ISM) z-density profile we used
 eq. (\ref{rhozp}) where: $\rho_{0}=n_{0}\mu m_{p}$ is the density in
 the disk plane; $m_{p}$ is the proton mass and $\mu $ is mean
 molecular weight for the disk (assumed to be 0.61). At 8 kpc we fix
 $n_{0}=1$. In Banerjee \& Jog (2008) it was shown that in the
 external regions of M31 the H values span the range between 300 and
 400 pc. Then we tested at 8 kpc 2 scale heights : H= 200 pc, and H=
 300 pc.

In our models we vary the number of SNeII in the OB association
(SNe). We consider four possible OB associations containing 50, 100,
250, 500 SNe respectively. Assuming an explosion energy of $10^{51}$
erg, the luminosities $L_0$ of these OB associations are 5 $\times
10^{37}$, $ 10^{38}$, 2.5 $\times 10^{38}$ and 5 $\times 10^{38}$ erg
s$^{-1}$ respectively. These numbers of massive stars in OB
associations are consistent with observations of Magnier et al. (1994)
who showed that the average number of massive stars in a OB
association in M31 is $\simeq$ 250.

\section{The Results for the galactic fountains in M31}

In this section we report the results concerning the fragmentation of
the superbubble, the formation of the cloud and the study of the
orbits of the galactic fountains in the cases of 8 kpc and 18 kpc. In
Figs. \ref{1} and \ref{2} we report our results at 8 kpc for OB
associations that can give rise to 250 SNeII, differing just in the
height scale of the ISM: H=200 pc and H=300 pc respectively. In
Fig. \ref{3} is considered the case of 250 SNeII at 18 kpc with H=300.
In Table 1 we report the ejection velocities for 50, 100, 250, and 500
SNeII at 8 kpc.

\begin{table}[htp] 
\caption{The ejection velocities of our clouds  as a function of the the number of SNeII and the height scale, in the case that the superbubble is fixed at 8 kpc.}
\label{ta4}
\begin{center}

\begin{tabular}{cccc}
  \hline\hline
\noalign{\smallskip}

 &$v_{o}$ [kms$^{-1}$]& $ v_{o}$ [kms$^{-1}$]  \\
\noalign{\smallskip}
 SNe& H=200  & H=300  \\
\noalign{\smallskip}
  \hline 
  50   & 37 & 28 \\
 100 &   46 &  35\\
 250 &   63 &  47\\
 500 &  79 & 60  \\

  \hline
 \end{tabular}
\end{center}
\end{table}

\subsection{$R_0=8$ kpc, H=200 pc}
\begin{figure}[!h]
\begin{center}
\includegraphics[width=0.50\textwidth]{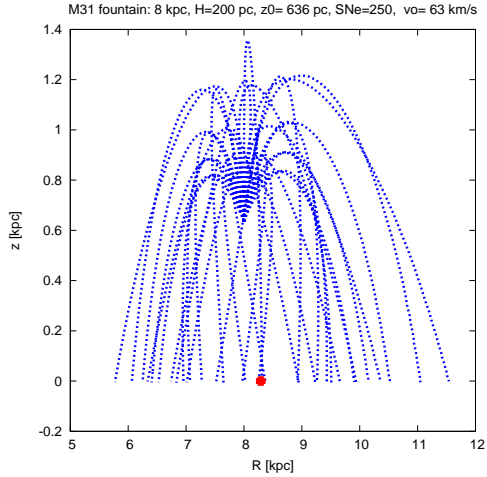} 
\caption{Galactic fountains reported in the meridional plane with the
  spatial initial conditions: $(R,z)$=(8 kpc, 636 pc). The red filled circle on the $R$ axis is the average falling radial coordinate.}
 \label{1}
\end{center}
\end{figure}

We computed the model with a OB association composed by 250 SNeII, at
8 kpc, and ISM height scale is 200 pc.  Using the Kompaneets
approximation we obtain that the superbubble is already fragmented in
clouds when its top side reaches $z_L=$ 636 pc ($\simeq$ 3H as found
in Spitoni et al. 2008, Mac Low 1994). This phase lasts 11.7 Myr.  In
Fig. \ref{1} we followed the orbit of our fountains. The initial
velocity in this model is 63 km/s and our main result is that the
clouds are generally thrown outward but the average landing coordinate
is 8.29 kpc (only $\Delta$R=0.29 kpc). The average orbit time is 42.7
Myr.

\subsection{$R_0=8$ kpc, H=300 pc}

\begin{figure}[!h]
\begin{center}
\includegraphics[width=0.50\textwidth]{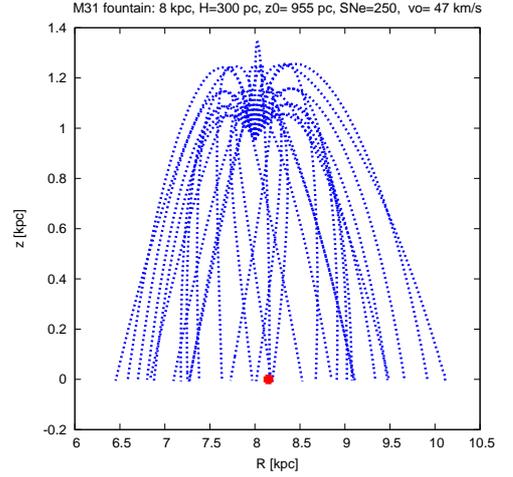} 
\caption{Galactic fountains reported in the meridional plane  with the spatial initial conditions: $(R,z)$=(8 kpc, 955 pc). The red filled circle on the $R$ axis is the average falling radial coordinate.}
 \label{2}
\end{center}
\end{figure}

Then we considered the model with the ISM medium height scale equal to
300 pc.  The superbubble is already fragmented in clouds when its top
side reaches $z_L=$ 955 pc ($\simeq$ 3H). This phase lasts 22.9 Myr.
In Fig. \ref{2} we followed the orbit of the fountains with an initial
velocity of 47 km/s and our main result is that the clouds are
generally thrown outward but the average landing coordinate is 8.15
kpc (only $\Delta$R=0.15 kpc). The average orbit time is 39.6 Myr.

\subsection{$R_0=18$ kpc, H=300 pc}

\begin{figure}[!h]
\begin{center}
\includegraphics[width=0.50\textwidth]{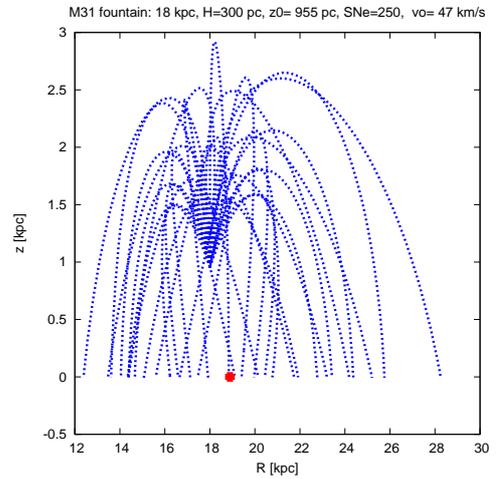} 
\caption{Galactic fountains reported in the meridional plane  with the spatial initial conditions: $(R,z)$=(18 kpc, 955 pc). The red filled circle on the $R$ axis is the average falling radial coordinate.
}
 \label{3}
\end{center}
\end{figure}

In Fig. \ref{3} we followed the orbit of fountains with an initial
velocity of 47 km/s and initial radial coordinate fixed at 18 kpc.
Also in this case the clouds are generally thrown outward and the
average landing coordinate is 18.89 kpc ($\Delta$R=0.89 kpc). The
average orbit time is 147.1 Myr.

Because of the fact that the average landing coordinate differs for the
throwing one for value always less than 1 kpc, we conclude that the
effects of galactic fountains cannot affect directly the chemical
evolution of M31, as we found for the Milky Way in Spitoni et
al. (2008).  Even if we consider the delay in the chemical evolution
enrichment, due to the fact that the clouds originated in the galactic
fountain processes take a finite time to orbit around the galaxy and
fall back onto disk, we find that this effect is negligible, since the
delays are the same as found in Spitoni et
al. (2010) for the Milky Way ($\sim$ 100  Myr). These delays have been proved not to have substantial effects on the process of chemical evolution. 

However, the infalling gas onto the disk, originated by the galactic
fountain, can be affected by the loss of angular momentum (Lacey
\& Fall 1985), in the same way as it happens for the primordial infall of gas. 
Therefore, this fountain gas  can
participate to the radial inflow of toward the galactic center, as we will
show in the next Sections.

 Concerning galactic fountains we also recall that in  the MaGICC (Brooke \& al. 
2012a,b,c  Pilkington \& al. 2012b)) program, where 
 hydrodynamical
  simulation in a cosmological context have been presented, fountains can be an important process  during  the galactic evolution.  
 
First of all, in Brook et al. (2012a) it was shown
that the majority of gas which loses angular momentum and falls into
the central region of the galaxy during the merging epoch is blown
back into the hot halo, with much of it returning later to form stars
in the disc. They proposed that this mechanism of redistribution of
angular momentum via a galactic fountain, can solve the angular
momentum/bulgeless disc problem of the cold dark matter paradigm.  It
was shown that this redistribution of angular momentum via large-scale
galactic fountains can lead to the formation of massive disc galaxies
which do not have classical bulges.

 In the MaGICC program was also found that a extensive and significant mixing of fountain material throughout the disk can have an important impact on the disk's chemistry. In fact, Brook
 et al. (2012c) showed that disc stars are dominated by smoothly
 accreted gas; a not insignificant amount of gas that feeds the thin
 disc does come from gas-rich mergers. Much of this is recycled to the
 disc via the hot halo, after being ejected from the star forming
 regions of the galaxy during starbursts. This large scale galactic
 fountain process allows the recycled gas to gain angular momentum
  and aids in the suppression of the ubiquitous G-dwarf problem
 (Pilkington et al.  2012b).

  The importance of strong feedback in
  securing correct scaling relations for disks and the correct coronal
  gas abundances in cosmological context has been discussed in Brook
  at al. (2012a, b, c) and Stinson et al. (2012).  It was shown that the
  scale of outflows invoked in models matches the observed absorption
  line features of local galaxies (Prochaska et al. 2011; Tumlinson et
  al. 2011).

\section{The chemical evolution model of M31} 

\subsection{Our reference model for M31} 
In order to reproduce the chemical evolution of the M31 disk, we started from
the one-infall chemical evolution model presented by
Marcon-Uchida et al. (2010), where the details can be found and then we introduced radial flows in the same way as in Spitoni \& Matteucci (2011) (see 4.1).
 In the starting
model, the galactic disk is divided into several concentric rings
which evolve independently without exchange of matter.
 
The disk is built up in an "inside-out" scenario which is a necessary 
condition to reproduce the radial abundance gradients when no radial gas flows are considered (Colavitti et al. 2008).  

For the star formation rate (SFR) a Schmidt law was used: 
 
\begin{equation} 
\Psi(r,t) = \nu \Sigma_{gas}^k(r,t) 
\end{equation}

The star formation efficiency $\nu$ is  assumed to vary with the
galactocentric distance in such a way:
\begin{equation}
\nu=24/R-1.5 \,\, Gyr^{-1} 
\end{equation}
 until it reached a minimum value of 0.5 Gyr$^{-1}$ and then is assumed to be constant.

This is suggested by the best model of Marcon-Uchida et al. (2010) in
order to reproduce the present day gas profile of M31. This gas
profile is different relative to that of the Milky Way: the gas
increases with decreasing galactic radius and, after reaching a peak
(at around 12 kpc) it decreases steeply towards the center, thus
suggesting a different scenario. This trend is probably the signature
of a very prominent spiral arm detectable in M31.
 
\subsection{The chemical evolution models of M31 in the literature}

 Several studies in the past have been addressed to the modelling of the chemical evolution of M31, here we compare our reference model of Marcon-Uchida et al. (2010) model of M31 with some of them. 

As done in Marcon-Uchida et al. (2010), the chemical evolution of M31
in comparison with that of the Milky Way has been discussed by Renda
et al. (2005) and Yin et al. (2009). Renda et al. (2005) concluded
that while the evolution of the Milky Way and M31 share similar
properties, differences in the formation history of these two galaxies
are required to explain the observations in detail. In particular,
they found that the observed higher metallicity in the M31 halo can be
explained by either  a higher halo star formation efficiency, or
 a larger reservoir of infalling halo gas with a longer halo
formation phase, which would lead to  younger stellar populations
in the M31 halo. Both pictures result in a more massive
stellar halo in M31, which suggests a possible correlation between the
halo metallicity and its stellar mass. Yin et al. (2009) concluded
that M31 must have been more active in the past than the Milky Way
although its current SFR is lower than in the Milky Way. They also
concluded that the star formation efficiency in M31 must have been
higher by a factor of two than in the Galaxy.  However, by adopting
the same SFR as in the Milky Way they failed in reproducing the
observed radial profile of the star formation and of the gas, and
suggested that possible dynamical interactions could explain these
distributions. 
The main
  difference between   the best model of M31 in Renda et al. (2005) and the model M31B of Marcon-Uchida et al. (2010), is the
  absence of any threshold in the star formation in the previous one.

In Yin et al. (2009) the infall prescription is the one presented by
Boissier \& Prantzos (2000), according to which the infall timescale
is assumed to be correlated with the flat rotational velocity for the
galaxy disk. In this work the star formation efficiency is
proportional to $\Sigma_{gas}\Omega$, where $\Omega$ is the rotation
speed of the gas, also in this model it was not consider a threshold in the star formation.

Carigi et a. (2012) presented a model of the chemical evolution of M31  computing also the Galactic Habitable Zones (GHZs) for this galaxy. We want to underline that they adopted the instantaneous recycling approximation, and therefore 
they can study only elements produced mainly by massive stars, such as oxygen.

\subsection{The implementation of  the  radial inflow  on  M31 chemical evolution model}

In Spitoni et al. (2011) we considered the case of radial inflow of gas 
for the Milky Way disk . Here we followed the same procedure to include 
radial gas flows in the disk of M31.
	  
We define the $k$-th shell in terms of the galactocentric radius $r_k$,
its inner and outer edge being labeled as $r_{k-\frac{1}{2}}$ and
$r_{k+\frac{1}{2}}$.  Through these edges, gas inflow occurs with velocity
v$_{k-\frac{1}{2}}$ and v$_{k+\frac{1}{2}}$, respectively. The flow
velocities are assumed to be positive outward and  negative inward.

The radial flow term to be added  into the chemical evolution equation is:

\begin{equation}
\left[ \frac{d}{dt} G_i(r_k,t) \right]_{rf} = -\, \beta_k \,
G_i(r_k,t) + \gamma_k \, G_i(r_{k+1},t),
\end{equation}
where $\beta_k$ and $\gamma_k  $ are, respectively:

\begin{equation}
\beta_k =  - \, \frac{2}{r_k + \frac {r_{k-1} + r_{k+1}}{2}}  
	\times \left[ v_{k-\frac{1}{2}}
	    \frac{r_{k-1}+r_k}{r_{k+1}-r_{k-1}}  \right]  
\end{equation}

\begin{equation}
\gamma_k =  - \frac{2}{r_k + \frac {r_{k-1} + r_{k+1}}{2}} 
	     \left[ v_{k+\frac{1}{2}}
	     \frac{r_k+r_{k+1}}{r_{k+1}-r_{k-1}} \right] 
	     \frac{\sigma_{A (k+1)}}{\sigma_{A k}},
\end{equation}
where $\sigma_{A (k+1)}$ and $\sigma_{A k}$ are the actual density
profile at the radius $r_{k+1}$ and $r_{k}$, respectively. 

\begin{figure}[!h]
\begin{center}
\includegraphics[width=0.50\textwidth]{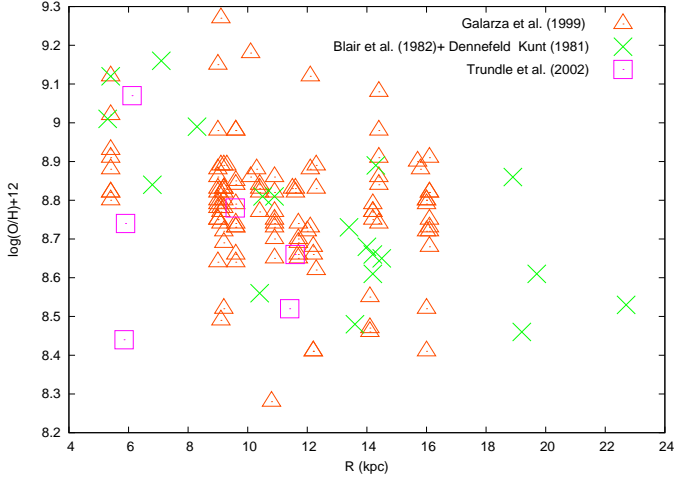} 
\caption{Oxygen abundances observed in the  HII Regions of M31. Data taken by: Galarza et al. (1999), Trundle et al. (2002), Blair
et al. (1982) and Dennefeld \& Kunth (1981).}
 \label{data}
\end{center}
\end{figure}

In our implementation of the radial inflow of gas in M31, only the gas
that resides inside the Galactic disk within the radius of 22 kpc can
move inward by radial inflow, as boundary condition we impose that
there is not flow of gas from regions outside the ring centered at 22
kpc.

\begin{table*}[htp]
\caption{Model parameters}
\scriptsize 
\label{models}
\begin{center}
\begin{tabular}{c|ccccc}
  \hline
\noalign{\smallskip}
\\
 Models & $\tau_d$& $\nu$&Threshold &Radial inflow  \\
  \\
\noalign{\smallskip}
\hline
\noalign{\smallskip}

M31B&   0.62 R (kpc) +1.62  Gyr & 24/R-1.5  Gyr$^{-1}$ &5 M$_{\odot}$pc$^{-2}$& no\\
\noalign{\smallskip}
\hline
\noalign{\smallskip}
M31N&   0.62 R (kpc) +1.62 Gyr & 2 Gyr$^{-1}$ &no&  no\\
\noalign{\smallskip}
\hline
\noalign{\smallskip}
M31R&   0.62 R (kpc) +1.62 Gyr & 2 Gyr$^{-1}$ &no& yes, variable speed\\
\noalign{\smallskip}
 \hline
\noalign{\smallskip}
\end{tabular}
\end{center}
\end{table*}

\section{Radial flows Results}

We have computed the oxygen gradient along the disk of M31 and compared to the available data. In particular, we adopted the same data as in Marcon-Uchida et al. (2010).
In Fig. \ref{data} we report the whole collection of the data
that we used in this paper: Galarza et al. (1999) (HII regions), Trundle et
al. (2002) (OB stars), Blair et al. (1982) and Dennefeld \& Kunth (1981) (supernova remnants and HII regions).

\begin{figure}[!h]
\begin{center}
\includegraphics[width=0.50\textwidth]{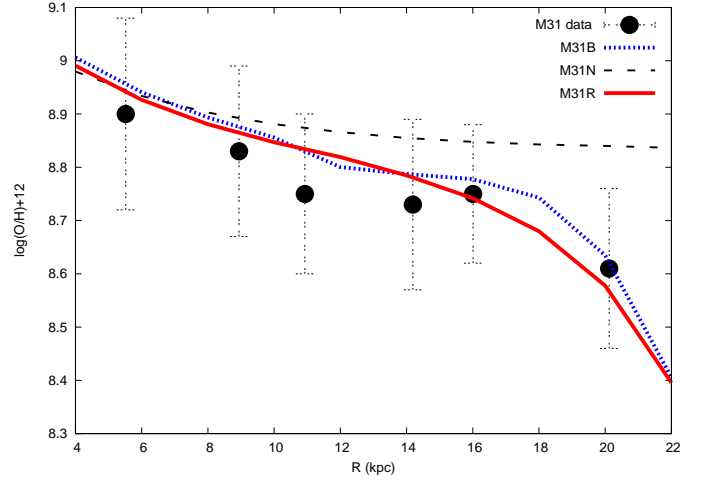} 

\caption{Radial oxygen abundance gradient. The blue dotted
line refers to the model M31B of Marcon-Uchida et al. (2010), the
black dashed line to the M31N model with a constant SF efficiency,
fixed at the value of 2 Gyr$^{-1}$, and without threshold,  the red solid line
  refers to the best fit model M31R with radial inflow of gas.
The filled circles and relatives error
bars are the observed values from HII regions.}
 \label{noth}
\end{center}
\end{figure}

 To better understand the trend of the data, we divided the data into
 six bins as functions of the galactocentric distance. In each bin, we
 computed the mean value and the standard deviations for the studied
 element, as reported with black filled circles and relatives errors
 in Fig. \ref{noth}. We are aware that there are some severe
 systematic uncertainties in each study that might justify larger
 error bars. In Fig. \ref{noth} we also plot the M31B model of
 Marcon-Uchida et al. (2010) without radial flows and we note that the
 model well fits the average trend of the data.  In Table
   \ref{models} we show the parameters of all the models we considered
   in this work: in particular, in column 2 there is the assumed
   inside-out law, in column 3 the assumed efficiency of star
   formation and in column 3 is indicated the presence or absence of
   radial flows. Model M31B is the best model adopted by Marcon-Uchida
   et al. (2010). This model contains an inside-out formation of the
   disk coupled with a variable efficiency of star formation and a
   threshold in the gas density, and it can reproduce the data without
   invoking radial flows.  However, the existence of a threshold in
   the star formation process has been questioned by several GALEX
   studies (e.g. Boissier et al. 2007) and without such a threshold
   the model would not so well reproduce the external parts of the
   disk of M31. In addition, the hypothesis of a variable star
   formation efficiency is not proven and with a constant star
   formation efficiency the gradient would look much flatter.
   Therefore, we can conclude as in Spitoni \& Matteucci (2011) did for 
the Milky Way disk, that
   radial gas flows can in principle be very important to reproduce
   the gradients along M31 disk, since all these processes
   are probably at work. To decide which of these
   processes is the most important in the formation of the disk, we
   will need more detailed data on the abundance, gas, and star
   formation rate gradients, as well as data on high redshift disks.

\begin{figure}[!h]
\begin{center}
\includegraphics[width=0.50\textwidth]{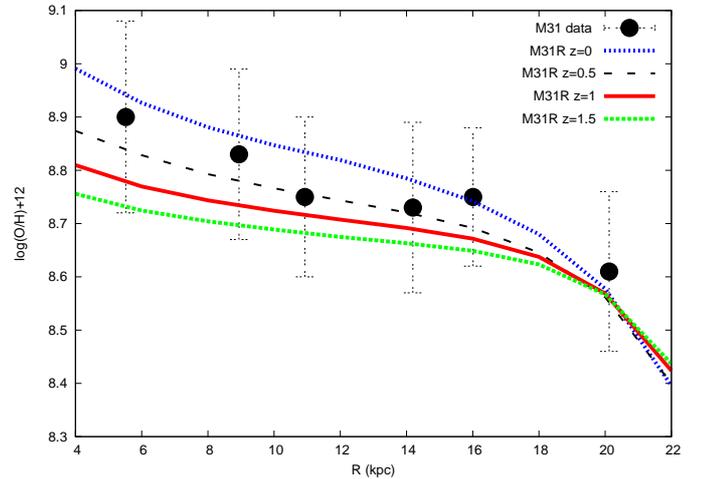} 

\caption{Evolution of the radial oxygen abundance gradient for the model M31R (the best model in this study) as a function of the redshifts z=0, 0.5, 1, 1.5.}
 \label{zM31R}
\end{center}
\end{figure}

\begin{figure}[!h]
\begin{center}
\includegraphics[width=0.50\textwidth]{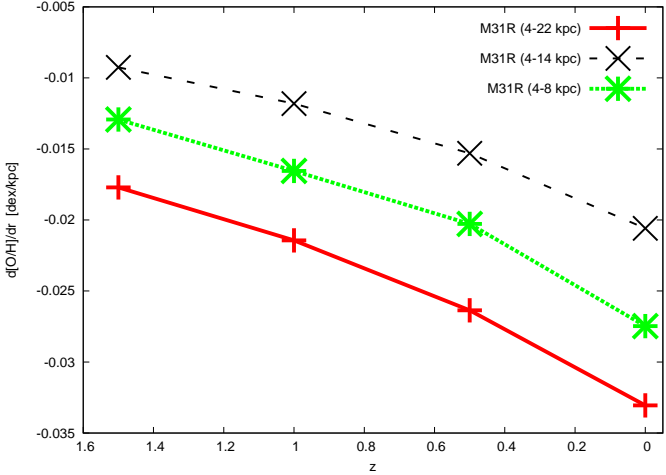} 

\caption{Here we plot the oxygen abundance gradient, d[O/H]/dr, computed
  between 4-8 kpc, 4-14 kpc, and 4-22 kpc for the model M31R (best model), as function of redshift.}
 \label{zM31BR8}
\end{center}
\end{figure}

 The first model we computed is M31N:  this model has a constant star
 formation efficiency, fixed at the value of 2 Gyr$^{-1}$, it does not
 assume a star formation threshold no radial flows. The abundance
 gradient obtained with this model is shown with the  dashed
 line in Fig. \ref{noth}. We see that this model fails to reproduce
 the gradient in the outer M31 regions.

   Then we assumed the same parameters as in M31N but we included the
   radial flows with a variable velocity: in particular, we used a
   linear relation between the radial inflow velocity and
   galactocentric distance, as done in Spitoni \& Matteucci (2011) for
   the disk of the Milky Way. In Fig.  \ref{noth} we label with the
   red solid line this model, the M31R model. The radial inflow
   velocity pattern requested to reproduce the data follows this
   linear relation:
   
\begin{equation}
|v_r|=0.05 R+0.45,
\end{equation}
and  spans the range of velocities between 1.55 and 0.65 km s$^{-1}$.

This model fits very well the O abundance gradient in the disk of M31, and we consider it our best model.

 In Fig. \ref{zM31R} we report the abundance gradient evolution
  for oxygen as a function of the redshift for the our best model M31R. In
  Pilkington et al. (2012a) it was shown the time evolution of the
  metallicity gradients $dZ/dr$ [dex/ kpc] using a suite of disk
  galaxy hydrodynamical simulations  and compared to the chemical evolution
  models of Chiappini et al. (2001) and Molla \& Diaz (2005).
   We computed the gradient d[O/H]/dR for the model M31R, at redshifts z$=0, 0.5, 1, 1.5$. In Fig. \ref{zM31BR8}  we show  the gradients for oxygen computed in the ranges 4-8 kpc,  4-14 kpc, and 4-22 kpc. We conclude that for the model M31R, where a variable gas inflow velocity is considered, a constant star formation efficiency, inside-out  formation and no threshold in the star formation are assumed, the abundance gradient steepens with time,  in accordance with the Chiappini et al. (2001) model. 
 
 The temporal evolution of the abundance gradients within cosmological
hydrodynamical simulations has been shown to be sensitive to the
spatial scale over which energy feedback operates (Pilkington et al
2012a).  Conventional feedback schemes with
"localized" energy feedback result in steep gradients at
high-redshift, which flatten with time towards redshift zero;
conversely, conventional schemes which distribute energy more
"globally" and/or enhanced feedback schemes which drive significant
outflows, circulation of the ISM, and radial gas flows, result in flat
gradients at high-redshift, which evolve little with time.  In this
sense, the models presented here are entirely consistent with the
latter enhanced feedback models (and vice versa).

\section{Conclusions}
In this paper we have studied the effects of galactic fountains and
radial inflows of gas on the predictions of a detailed chemical
evolution model for M31.  Our main conclusions can be summarized as
follows:

\begin{itemize}

\item Considering the average number of massive stars in a OB
  association in M31, the range of the cloud orbits is quite
  small. The clouds are generally directed outwards but the average
  landing coordinates differ from the throwing coordinates by values
  less than 1 kpc. Because of this fact, we conclude that galactic
  fountains cannot affect the chemical evolution of M31, as we found
  for the Milky Way in Spitoni et al. (2008).  The reason for that is
  that inside 1 kpc, such as in the solar neighborhood, the gas is
  well mixed.

\item The average time delay produced by a galactic fountain generated by an
  OB association in M31 is $\simeq$  100 Myr. As we suggested for
  the Milky Way, in Spitoni et al. (2009), such a small delay time 
  has a negligible
  effect on the abundance gradients in the Galactic  disk.

\item  We started by adopting the best chemical evolution model of
  Marcon-Uchida et al. (2010) for the disk of M31. This model, in
  order to reproduce the observed abundance gradient, needs to assume
  a threshold in the star formation and an inside-out formation plus a
  star formation efficiency varying with the galactocentric distance.
  However, the existence of a threshold in the star formation has been
  questioned and the variable efficiency of star formation with
  galactocentric distance is not strongly physically motivated. On the
  other hand, a moderate inside-out formation for galactic disks seems
  to be observed at high redshift (Munoz-Mateos et al. 2007).

\item The radial gas flow velocity, that we have found to be most
  consistent with the data,  varies linearly with the galactocentric
  distance and spans a range between 0.65 and 1.55 km s$^{-1}$. This conclusion
  holds for the M31 chemical evolution  model with inside-out formation, 
  without 
  threshold, 
  and a constant star formation efficiency fixed at 2 Gyr$^{-1}$.

\item  We showed that the abundance gradient d(O/H)/dr  steepens with  time for our best model which assumes no gas threshold in the star formation, an inside-out formation of the disk, a constant star formation efficiency along the disk and radial gas flows.

\item  Finally, we conclude, in agreement with Spitoni \& Matteucci
  (2011), that  also for M31 the radial gas flows can be very 
  important to reproduce the gradients along the disk, although an
  inside-out formation coupled with a variable efficiency of star
  formation and threshold in gas density can also closely reproduce
  the data without radial flows. To decide which of these processes
  are the most relevant in the formation of the disk, we will need
  more detailed  and precise data on the abundance, gas, and star formation 
 rate gradients in M31, as well as more data on high redshift disks.

\end{itemize}

\begin{acknowledgements}
 We thank the referee B. K. Gibson for his suggestions which improved the paper.
E. Spitoni acknowledges financial support from the FCT by grant SFRH/BPD/78953/2011. F. Matteucci aknowledges financial support from PRIN MIUR 2010-2011, project ``The
Chemical and dynamical Evolution of the Milky Way and Local Group
Galaxies'', prot. 2010LY5N2T.
M. M. Marcon-Uchida acknowledges financial support from FAPESP (2010/17142-4).

\end{acknowledgements}

\end{document}